\documentclass[%
reprint,
showkeys,
superscriptaddress,
nofootinbib,
nobibnotes,
amsmath,amssymb,
prstper,
floatfix,
]{revtex4-2}

\usepackage{graphicx}
\usepackage{dcolumn}
\usepackage{bm}
\usepackage{hyperref}
\usepackage[mathlines]{lineno}
\usepackage{xcolor}
\usepackage{mathtools}

\hypersetup{colorlinks=true,%
            linkbordercolor=teal,
            citecolor=teal,
            urlcolor=teal,
            filecolor=teal,
            linkcolor=teal}

\def\micrometer{$\mathrm{\text\textmu}$m}

\begin{document}


\title{Fourier Transform-Based Post-Processing Drift Compensation and Calibration Method for Scanning Probe Microscopy}


\author{M. Le Ster}
\email{maxime.lester@fis.uni.lodz.pl}

\author{S. Paw{\l}owski}

\author{I. Lutsyk}

\author{P. J. Kowalczyk}
\email{pawel.kowalczyk@uni.lodz.pl}

\affiliation{University of Lodz, Faculty of Physics and Applied Informatics, Department of Solid-State Physics, Pomorska 149/153, Lodz, 90-236, Poland}

\date{\today}

\begin{abstract}
Scanning probe microscopy (SPM) is ubiquitous in nanoscale science allowing the observation of features in real space down to the angstrom resolution. The scanning nature of SPM, wherein a sharp tip rasters the surface during which a physical setpoint is maintained via a control feedback loop, often implies that the image is subject to drift effects, leading to distortion of the resulting image. While there are \emph{in-operando} methods to compensate for the drift, correcting the residual linear drift in obtained images is often neglected. In this paper, we present a reciprocal space-based technique to compensate the linear drift in atomically-resolved scanning probe microscopy images without distinction of the fast and slow scanning directions; furthermore this method does not require the set of SPM images obtained for the different scanning directions. Instead, the compensation is made possible by the a priori knowledge of the lattice parameters. The method can also be used to characterize and calibrate the SPM instrument.\end{abstract}


\maketitle
\newpage

\section{Introduction}

Scanning probe microscopy (SPM) has revolutionized nanotechnology and our understanding of nature \cite{Bian2021}, allowing for the real space observation and characterization of nano-sized objects (and under certain conditions, local manipulation), such as individual atoms \cite{Natterer2017}, molecules \cite{Hamalainen2014}, nanoparticles \cite{Kano2015}, surfaces of crystalline bulk materials \cite{Binnig1983}, 1D crystals such as nanotubes \cite{Bachtold2000}, as well as larger macromolecules such as DNA \cite{Hansma2001} and viruses \cite{Kuznetsov2011}, and even larger objects of biological nature \cite{Allison2010}. Among the many SPM techniques, the most popular today are scanning tunneling microscopy (STM) \cite{Binnig1983, Binnig1987}, wherein a biased atomically-sharp metallic probe is brought near the surface of the sample maintaining a quantum tunneling current setpoint; and atomic force microscopy (AFM) \cite{Binnig1986} where the deflection of the cantilever (attached to the sharp tip-probe), in contact or near-contact with the surface, is measured via optical methods. Beyond topography-based SPM (yielding height maps), there are many other SPM techniques tangential to STM and AFM dedicated to additional electronic characterizations that are worth mentioning such as scanning tunneling spectroscopy \cite{Feenstra1994} (mapping the electronic density of states), Kelvin-probe force microscopy \cite{Melitz2011} (mapping the electronic work function) and magnetic force microscopy \cite{Hartmann1999} (sensitive to the surface magnetization).

The inherent scanning nature of SPM mean that the image acquisition is simultaneously obtained with the rastering of a piezoelectric tube carrying the tip-sensor. While this has the advantage of achieving extremely high resolution images (with lateral resolutions down to the angstrom scale, or even below \cite{Giessibl2019}) and the ability to characterize local physical properties, SPM often suffers from drift effects of various origins, such as piezoelectric hysteresis and scanner creep \cite{Meyer2004} (due to mechanical relaxation of the piezoelectric components) and thermal drift \cite{Marinello2011} (due to differing thermal expansion coefficients across different constituent parts of the SPM instrument, whether in the sample stage or in the scanning stage elements), inducing image distortion.

While there exist \emph{in-operando} methods to compensate the drift effects (\emph{e.g.}, relying on the monitoring a feature in two SPM images obtained sequentially \cite{Clifford2009}), it is possible that distortion still occurs in the final image due to subsequent changes in the drift velocity vector, its typical values being in the nm/s range at room temperature \cite{Meyer2004}. A number of post-processing methods (\emph{i.e.}, after image acquisition) to compensate for thermal drift have previously been developed using a number of different approaches; Fourier-transform cross-correlation \cite{Mantooth2002}, phase correlation \cite{Yang2008}, or relying on images obtained in different scanning directions \cite{Lapshin2007}, and lastly addressing non-linear distortion effects \cite{Clayton2005}. Another possible scenario where post-processing correction is required occurs when sudden changes in the tip-sample interactions lead to undesirable effects such as an unforeseen loss of sharpness or a substantial noise increase, before \emph{in-operando} drift compensation methods are applied. For this reason it is often desirable to have post-processing drift compensation methods in order to ensure proper mapping, \emph{i.e.}, spatial integrity of the final image.

In this paper, we present a technique based on the Fourier transform of the acquired SPM image that allows for post-processing compensation of the linear drift which may exist in images whether or not \emph{in-operando} efforts have been undertaken to curb drift effects. The reciprocal space nature of the technique requires that the real space images contains periodic features in order to easily correct the skewed raw image. A similar method has been developed previously \cite{Aketagawa1995}, however its implementation requires several additional steps (such as the determination of the drift velocity vector) and the knowledge of the fast and slow scanning directions, irrelevant in our method. Another strength of the method is that it can be applied on a single image (without requiring the different images obtained for different scanning directions) obtained using an SPM instrument that has not been calibrated along the $(x,y)$ scanning directions. After introducing the model and its implementation, we benchmark and illustrate the method on two images that contain periodic modulations: \emph{(i)} an atomically-resolved graphite sample imaged by STM and \emph{(ii)} a calibration grating sample imaged by AFM. 

\section{Model and Implementation}

\begin{figure}[t] 
    \centering
	\includegraphics[width=\columnwidth, trim={0.4cm 0.4cm 0.4cm 0.4cm}]{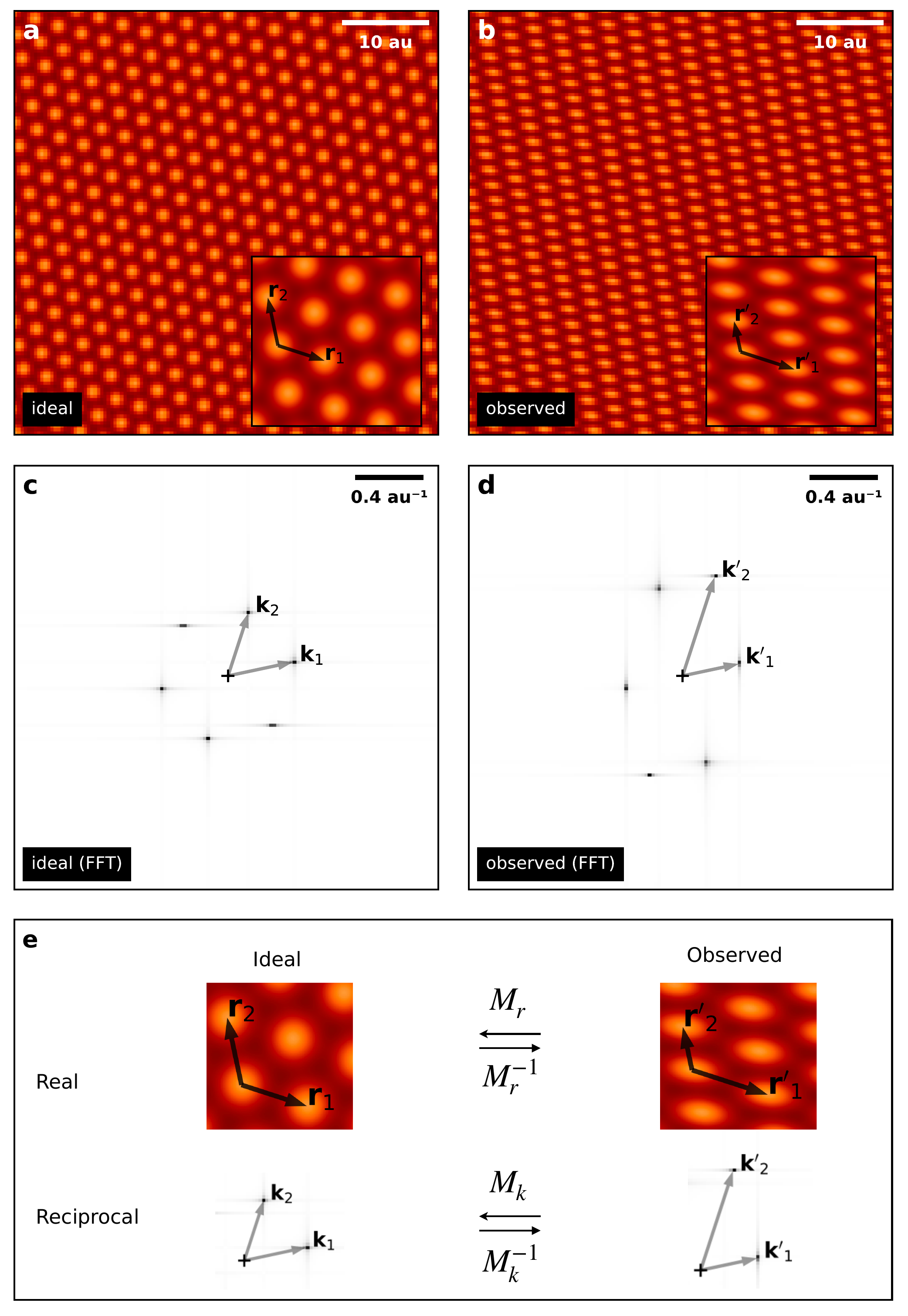}
    \caption{(a) Ideal SPM image of a triangular lattice (in arbitrary units of spatial dimensions). (b) Drift-containing SPM image of a triangular lattice. (c) FFT of the ideal SPM image in (a). (d) FFT of the drift-containing SPM image in (b). (e) Linear transformations between the different vector bases. Insets in (a, b) are higher resolution cropped images showing the ideal ($\mathbf{r}_1, \mathbf{r}_2$) and observed ($\mathbf{r}'_1, \mathbf{r}'_2$) primitive real space vectors, respectively. Arrows in (c, d) correspond to the primitive reciprocal lattice vectors ($\mathbf{k}_1, \mathbf{k}_2$) and ($\mathbf{k}'_1, \mathbf{k}'_2$) for the ideal and observed images, respectively.}
    \label{fig:1}
\end{figure}

Figure~\ref{fig:1}a shows a simulated ideal drift-free SPM image consisting of a triangular lattice (see inset for real space base vectors), which can be of various origins: primitive unit cell at the surface of a crystalline material (bulk or 2D material, typical scale: $\sim3~\mathrm\AA$), superlattice (such as moir\'e patterns \cite{LeSter2019}, surface reconstruction \cite{Binnig1987}, or charge-density-wave, \emph{e.g.}, in 1T-TaS$_2$ \cite{Lutsyk2023}, typical scale: $\sim10-100~\mathrm{\AA}$), or due to artificial nano-patterning (\emph{e.g.}, via lithography, typical scale: $10-10000$~nm). Figure~\ref{fig:1}b shows an equivalent SPM image, this time obtained under significant linear drift conditions, resulting in a strongly distorted image (compare insets in Figs.~\ref{fig:1}a and~\ref{fig:1}b to visualize distortion of the primitive real space vectors). The unit cell vectors $(\mathbf{r}_1, \mathbf{r}_2)$ are known \emph{a priori} by means of theory or other experimental methods; alternatively the sample may be a calibration sample.

Figures~\ref{fig:1}c and~\ref{fig:1}d are the fast Fourier transforms (FFTs) of the ideal and drifted real space images in Figs.~\ref{fig:1}a and~\ref{fig:1}b respectively. The sharp features resolved in FFTs correspond to the primitive reciprocal lattice vectors $\mathbf{k}_1$ and $\mathbf{k}_2$ ($\mathbf{k}'_1$ and $\mathbf{k}'_2$). 

\subsection{Matrix definitions}

The essence of the model is to obtain the linear matrix $M_r$ that linearly transforms the observed lattice ($\mathbf{r}'_1, \mathbf{r}'_2$) into the ideal lattice ($\mathbf{r}_1, \mathbf{r}_2$) which would be resolved in a drift-free imaging conditions on a calibrated instrument. The distorted image (under the $M_r$ linear transformation) will then display a corrected lattice $(\mathbf{r}_1, \mathbf{r}_2)$. The matrix scheme associating the different vector bases is detailed in Fig.~\ref{fig:1}e. The matrix $M_r$ transforms ($\mathbf{r}'_1, \mathbf{r}'_2$) into ($\mathbf{r}_1, \mathbf{r}_2$), and the reciprocal space equivalent $M_k$ transforms ($\mathbf{k}'_1, \mathbf{k}'_2$) into ($\mathbf{k}_1, \mathbf{k}_2$). The inverse matrices $M_r^{-1}$ and $M_k^{-1}$ are the inverse operations, transforming $(\mathbf{r}_1, \mathbf{r}_1)$ into $(\mathbf{r}'_1, \mathbf{r}'_1)$ and $(\mathbf{k}_1, \mathbf{k}_1)$ into $(\mathbf{k}'_1, \mathbf{k}'_1)$, respectively.

Instead of directly computing the real space matrix $M_r$, the inputs in the method are the observed and ideal \emph{reciprocal} lattice parameters, in order to obtain a reciprocal space matrix $M_k$. The real space matrix $M_r$ can then be obtained by matrix inversion and transposition. The reason this is that the observed reciprocal lattice parameters $(\mathbf{k}'_1, \mathbf{k}'_2)$ are easy to extract from the raw data (\emph{i.e.} by simple observation of the FFT) as opposed to the real space vectors $(\mathbf{r}'_1, \mathbf{r}'_2)$ whose characterization is subject to a higher uncertainty and possibly local variation across the same image. The ideal reciprocal lattice vectors $\mathbf{k}_1$, $\mathbf{k}_2$ can be calculated using standard formulae \cite{Kittel}.

Solving for a general matrix $M_k$ such that $\mathbf{k}_1 = M_k \times \mathbf{k}'_1$ and $\mathbf{k}_2 = M_k \times \mathbf{k}'_2$, we can obtain the reciprocal space matrix $M_k$:
\begin{equation}
\label{eq:mk_general}
\begin{split}
M_k &= \frac{1}{\Delta}\begin{pmatrix}
k_{1x} k'_{2y} - k_{2x} k'_{1y} & k_{2x} k'_{1x} - k_{1x}k'_{2x} \\
k_{1y} k'_{2y} - k_{2y} k'_{1y} & k_{2y} k'_{1x} - k_{1y}k'_{2x}
\end{pmatrix},\\
\Delta &= k'_{1x}k'_{2y} - k'_{2x}k'_{1y}.
\end{split}
\end{equation}

While a correction of the reciprocal space data (\emph{i.e.}, FFT) can be achieved using $M_k$, the motivation concerns the drift correction of \emph{real space} images, and we are now concerned with the matrix $M_r$. It can be shown that the matrix $M_r$ is derived from $M_k$ as follows:
\begin{equation}
\label{eq:inversetranspose}
M_r = (M_k^T)^{-1}
\end{equation}
where $M^T$ denotes the transpose of $M$ (see supplementary information for a formal demonstration).

\subsection{Implementation}

We briefly describe the implementation of the model using the {\fontfamily{lmtt}\selectfont warpAffine} function from the {\fontfamily{lmtt}\selectfont openCV} library \cite{OpenCV} to perform the linear transformation of the raw SPM image. The {\fontfamily{lmtt}\selectfont warpAffine} function returns the straightened image (with no dimension metadata) and takes the following inputs: $(a)$ the image array ($n'\times m'$ pixels, no specified pixel or image dimensions), $(b)$ an affine $(2\times3)$ matrix (of which the third column is a translation vector in pixel coordinates), $(c)$ the final image size (in pixels) and $(d)$ an optional border mode. The four inputs as well as the final image dimensions are described in the following paragraphs.

\paragraph{Image array.} We operate primarily on the $(n'\times m')$ real space image array obtained by the SPM instrument (we also operate on the FFT image later).

\paragraph{$(2\times3)$ matrix.} The input matrix $M$ is a $(2\times3)$ matrix representing an affine transform, \emph{i.e.}, $M \times X = M'\times X+ T$ with $M'$ the $(2\times2)$ linear matrix, $X$ the vector to modify and $T$ a translation vector. We find that it is convenient to use a linear $(2\times2)$ matrix that is \emph{normalized}, which we define in the following. Any linear matrix $M$ can be thought of as a combination of a biaxial scaling matrix $M_s$, a rotation matrix $M_\theta$ and a shear matrix $M_\tau$ as follows:
\begin{equation}
\label{eq:matrixcomponents}
\begin{split}
M &= M_s\times M_{\theta} \times M_{\tau}\\
M_s &= \begin{pmatrix}
s_x & 0\\
0 & s_y
\end{pmatrix}\\
M_\theta &= \begin{pmatrix}
\cos\theta & -\sin\theta\\
\sin\theta & \cos\theta
\end{pmatrix}\\
M_\tau &= \begin{pmatrix}
1 & \tau\\
0 & 1
\end{pmatrix}
\end{split}
\end{equation}
where $s_x$ and $s_y$ are scaling factors along the $x$ and $y$ directions, $\theta$ is the rotation angle and $\tau$ the shear parameter. The order of the matrix multiplication is not crucial (one could alternatively scale, then shear, and finally rotate; or in another order), however it is important to keep the order of operation consistent since matrix multiplication is not commutative. The \emph{normalized} matrix $\tilde{M}$ is then:
\begin{equation}
\tilde{M} = M_\theta \times M_\tau.
\end{equation}

It is worth noting that $\det(\tilde{M})=1$, \emph{i.e.}, a linear operation under a normalized matrix preserves areas. The $(2\times3)$ is then a composite matrix made of the $(2\times2)$ \emph{normalized} matrix $\tilde{M_r}$, and of a $(1\times2)$ translation vector $T$ which we define in the following. The translation vector $T$ ensures that that every pixel coordinates $(p,q)$ of the final image are such that $p,q\geq0$, and that $(0,0)$ is a corner of the output image. With \mbox{$C = \{{0 \choose 0}, {n'-1 \choose 0}, {0\choose m'-1}, {n'-1 \choose m'-1}\}$} the set of corner pixel coordinates of the raw image, the translation vector $T$ is then:
\begin{equation}
\label{eq:translationvector}
T=-\begin{pmatrix}
\min{\left\{ \left( \tilde{M}_r \times C_i \right) \cdot {1\choose0}\right\} }\\
\min{\left\{ \left( \tilde{M}_r \times C_i \right) \cdot {0\choose1}\right\} }
\end{pmatrix}
\end{equation}

Given a non-normalized matrix $M=\begin{pmatrix}a&b\\c&d\end{pmatrix}$, its rotation, shear and scaling parameters defined in eq. (\ref{eq:matrixcomponents}) can be retrieved:
\begin{equation}
\label{eq:matrixcomponents1}
\begin{split}
\theta &= \mathrm{atan2}(c, d)\\
\tau &= \frac{a\sin\theta + b\cos\theta}{a\cos\theta - b\sin\theta}\\
s_x &= \frac{a}{\cos\theta + \tau\sin\theta} = \frac{b}{\tau\cos\theta-\sin\theta}\\
s_y &= \frac{c}{\sin\theta} = \frac{d}{\cos\theta}
\end{split}
\end{equation}
with $\mathrm{atan2}$ the $2$-argument arctangent function. If $s_x$ or $s_y$ diverge using the left-hand equalities ($\cos\theta+\tau\sin\theta=0$ or $\sin\theta=0$), the right-hand expression may be used. The normalization of a matrix $M$ is achieved by first extracting its $\theta$ and $\tau$ components with eqs. (\ref{eq:matrixcomponents1}) and using eq. (\ref{eq:matrixcomponents}) with $s_x=s_y=1$.

\paragraph{Output dimensions.} The dimensions of the output image $(n, m)$ can be obtained by considering the difference between the maximum and minimum pixel coordinates (for both column and row indices) upon transformation of the corner pixel coordinates with $\tilde{M}_r$. The updated dimensions (in pixel units) are then:

\begin{equation}
\label{eq:pixels}
\begin{split}
\begin{pmatrix}
n\\
m
\end{pmatrix}=
&\begin{pmatrix}
\max{\left\{ \left( \tilde{M}_r \times C_i \right) \cdot {1\choose0}\right\}} \\
\max{\left\{ \left( \tilde{M}_r \times C_i \right) \cdot {0\choose1}\right\}}
\end{pmatrix}+T\\
\end{split}
\end{equation}
The dimensions (in pixel units) are then rounded to the closest integer number.

\paragraph{Border mode.} An image that is warped such that $\tau\neq0$ (\emph{i.e.}, that undergoes shear) will not be rectangular; rather, its corners will describe a parallelogram. The border mode argument set to {\fontfamily{lmtt}\selectfont BORDER\_CONSTANT} will fill the values of the undefined pixels in the border. To minimize artifacts in the FFT of the image (due to the presence of discontinuity at the border), it is desirable to set the value to the average value of the SPM image.

\paragraph{Spatial dimensions.}
Finally, aside from the {\fontfamily{lmtt}\selectfont warpAffine} operation (which returns a dimensionless image), it is important to consider the \emph{spatial} dimensions of the image to properly assign the dimensions of the output image, crucial for accurate compensation. Using an approach identical to the dimensions in pixel space, where \mbox{$D=\{{0\choose0},{L'_x\choose0},{0\choose L'_y},{L'_x \choose L'_y}\}$} the set of corner spatial coordinates, the dimensions $L_x$, $L_y$ of the final image after the warping procedure are:
\begin{equation}
\label{eq:dimensions}
\begin{split}
\begin{pmatrix}
L_x\\
L_y
\end{pmatrix}=
&\begin{pmatrix}
\max{\left\{ \left( M_r \times D_i \right) \cdot {1\choose0}\right\}} \\
\max{\left\{ \left( M_r \times D_i \right) \cdot {0\choose1}\right\}}
\end{pmatrix}\\
-&\begin{pmatrix}
\min{\left\{ \left( M_r \times D_i \right) \cdot {1\choose0}\right\}} \\
\min{\left\{ \left( M_r \times D_i \right) \cdot {0\choose1}\right\}}
\end{pmatrix}
\end{split}
\end{equation}

\section{Results}

\subsection{Scanning Tunneling Microscopy}
\label{section:stm}

\begin{figure}[t] 
    \centering
	\includegraphics[width=\columnwidth, trim={0.3cm 0.8cm 0.5cm 0.3cm}]{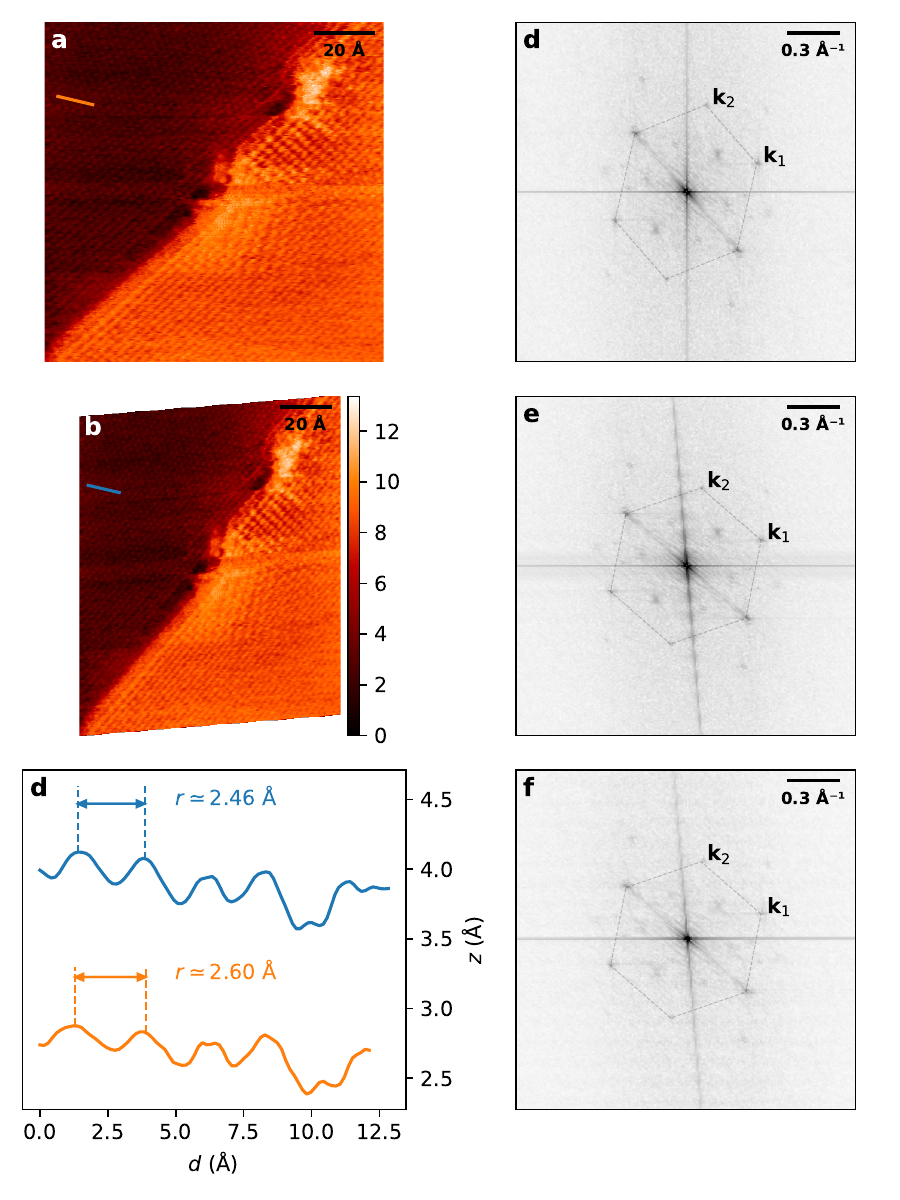}
    \caption{(a) Raw STM image of HOPG, at the location of a step edge (diagonal line). (b) Drift-corrected STM image in (a) (color scale in \AA). (c) Line profiles extracted from lines in (a) and (b) of the same colors (vertically offset for clarity). (d) FFT of the raw data in (a). (e) FFT of the corrected data in (b). (f) Warped FFT of the raw data in (d). The hexagons in FFTs (d-f) connecting the reciprocal lattice points of graphite allow to visualize the presence of drift in (d) and its compensation in (e, f).}
    \label{fig:2}
\end{figure}

We illustrate the model using an experimental STM image shown in Fig.~\ref{fig:2}a. The two main areas (dark brown and orange) in the image correspond to different atomically flat planes of the highly-ordered pyrolithic graphite (HOPG) crystal (space group: $P6_3/mmc$). The two areas of the image are offset vertically at the step edge by $\sim6$~\AA, \emph{i.e.}, 2 atomic planes parallel to the $c$ axis.  The corrugation on the top-left area of the image corresponds to the unit cell of graphite, of well-known lattice constants ($r_1=r_2=2.46~\mathrm{\AA}$, forming a triangular lattice along the (0001) plane \cite{Yang2018}). This SPM image may be relevant for the investigation of the physics related to the presence of defects along the edge of the atomic terrace (see \emph{e.g.}, the brighter features located at kinks and the standing waves near the edge of the top terrace). Compensating for the drift in such image may be of significant importance for correct real space characterization.

The line profile, measured from the orange line in Fig.~\ref{fig:2}a, shown in Fig.~\ref{fig:2}c (with the same color) indicates a lattice parameter of $\simeq 2.60$~\AA~along this particular direction, suggesting either a calibration error (related to the piezoelectric tube which drives the tip motion across the surface) or the presence of thermal drift. The FFT of the raw data is shown in Fig.~\ref{fig:2}d. The primitive reciprocal lattice vectors $\mathbf{k}'_1$ and $\mathbf{k}'_2$ are clearly resolved (as well as the composite reciprocal lattice vector $\mathbf{k}_1-\mathbf{k}_2$), and it is clear that they do not form a regular hexagon (see dashed lines), which is to be expected in an SPM image suffering from linear drift.

Knowing the expected lattice parameters $(\mathbf{r}_1, \mathbf{r}_2)$, the ideal primitive reciprocal vectors $(\mathbf{k}_1, \mathbf{k}_2)$ are calculated; the observed primitive reciprocal vectors $(\mathbf{k}'_1, \mathbf{k}'_2)$ are obtained from FFT in Fig.~\ref{fig:2}b. The reciprocal matrix $M_k$ is then easily obtained using eq. (\ref{eq:mk_general}), the real space matrix $M_r$ is computed using eq. (\ref{eq:inversetranspose}). The normalized matrix $\tilde{M_r}$, the translation component $T$ from eq. (\ref{eq:translationvector}), and finally the expected dimensions in pixel and spatial coordinates eqs. (\ref{eq:pixels}, \ref{eq:dimensions}) are also computed. The method is then applied on the raw STM image in Fig.~\ref{fig:2}a, obtaining the image in Fig.~\ref{fig:2}b. The line profile extracted from the same location as previously (now in blue) is shown in Fig.~\ref{fig:2}c; the apparent lattice parameter is now in agreement with the theoretical value of graphite, confirming the success of the method. Note that the corrected image is skewed, resulting from a non-zero shear ($\tau=0.066$). The other real space linear transformation parameters are $s_x=0.914$, $s_y=1.118$ and $\theta=3.84^\circ$. Figure~\ref{fig:2}e shows the FFT of the drift-corrected image of Fig.~\ref{fig:2}b (where the border of Fig.~\ref{fig:2}b is filled with $0$ values). The sharp features in the FFT corresponding to the primitive reciprocal lattice now describe a regular hexagon, further confirming a satisfactory drift compensation.

Figure~\ref{fig:2}f is obtained by applying the {\fontfamily{lmtt}\selectfont warpAffine} operation on the FFT image in Fig.~\ref{fig:2}d using a normalized $M_k$ matrix instead. There are no structural differences to be observed with Fig.~\ref{fig:2}e; however the artifacts present in the FFT differ slightly. This is particularly evident in case of large shear values, where significant edge effects (due to the border populated with $0$ values) can be observed in the FFT of the skewed image. In this example however, this effect is relatively weak, and either FFT images can be used for further analyses if required.


\subsection{Atomic Force Microscopy}
\label{section:afm}

\begin{figure}[t]
    \centering
	\includegraphics[width=\columnwidth, trim={0.9cm 0 0.9cm 0}]{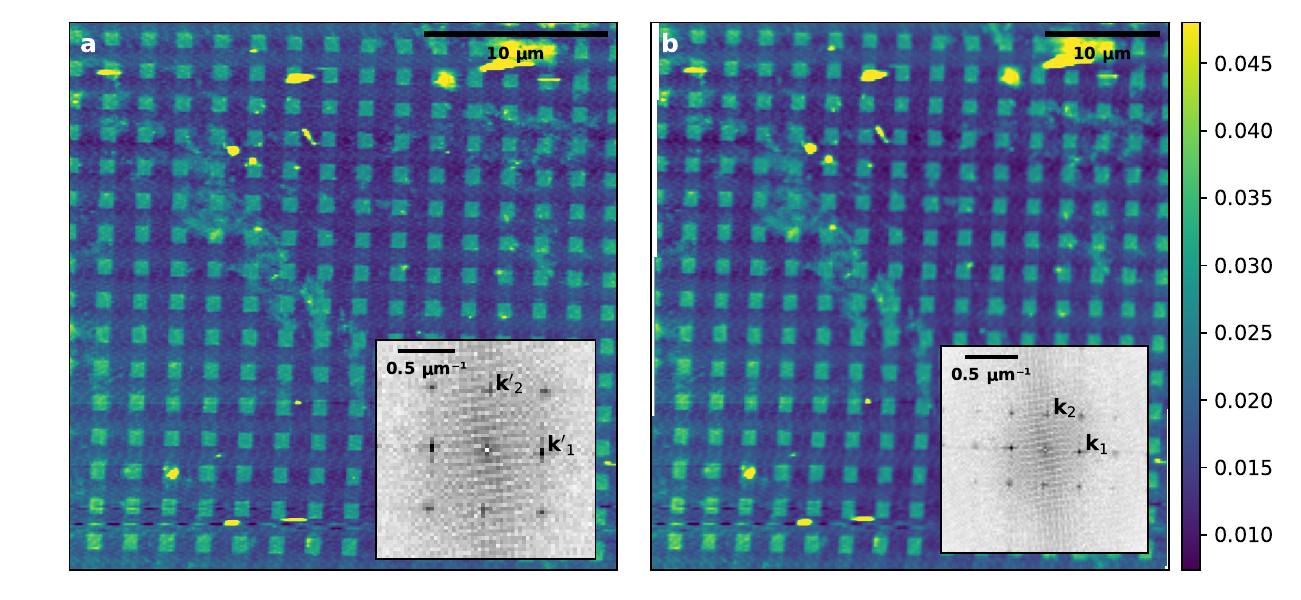}
    \caption{(a) Raw AFM image of a nano-patterned Si sample. (b) Corrected AFM image (color scale in \micrometer). Insets are the FFT of the images shown in main panels.}
    \label{fig:3}
\end{figure}

We now focus on an AFM image of a commercially available SiO$_2$/Si calibration grating sample (TGQ1 from NT-MDT), shown in Fig.~\ref{fig:3}a. The sample is characterized by the presence of SiO$_2$ pillars with bare Si material in between them (dark blue). In contrast to the previous case, the topography modulation in the SPM image here results from the artificial periodic pillar structure. The nominal distance between the pillars, arranged in a square lattice, is $3.00\pm0.05$~\micrometer. Quick inspection of the raw AFM image indicates a scaling error; the pillars appear to be separated by a smaller apparent distance, \emph{i.e.}, $\simeq2$~\micrometer~in both directions. The FFT of the raw image (see inset in Fig.~\ref{fig:3}a) further confirms this in reciprocal space, where the two primitive reciprocal vectors $\mathbf{k}'_1$ and $\mathbf{k}'_2$ associated with the periodic pillar structure are arranged $\sim0.5$~\micrometer$^{-1}$ from the origin. This is consistent with a significant calibration error, which we correct using the method, as shown in the following.

The as-measured primitive reciprocal coordinates (from the FFT in the inset of Fig.~\ref{fig:3}a) and the nominal real lattice parameters ($3.0\times3.0$~\micrometer$^{2}$ in a square lattice) are input into the model. Figure~\ref{fig:3}b shows the AFM image after correction. The image dimensions is now consistent with the nominal periodic spacing. The associated transformation parameters are $s_x = 1.509$, $s_y=1.613$, $\tau=0.017$ and $\theta=0.17^\circ$. The non-zero shear parameter $\tau$ indicates the presence of trace drift in the AFM image (albeit about 5 times weaker than in the first example in Fig.~\ref{fig:2}). The scaling factors $s_x$ and $s_y$ can then be used to calibrate the scanning stage of the AFM instrument.

\section{Discussion and conclusion}

As shown in the two examples above, the FFT-based distortion method reliably corrects for shear (typically resulting from drift) and scaling errors (resulting from improper or absent calibration of the SPM scanning stage). The method allows to retrieve the different scaling, rotation and shear parameters contributing to a general linear transformation. The correction factors $s_x$ and $s_y$ indicate how to scale the $x$ and $y$ directions of the piezo-electric scanning stage appropriately in order to obtain a 1:1 scale correspondence between the image and the object.

We believe that our method and the simplicity of its implementation should encourage scientists using SPM imaging for crystalline or patterned samples to correctly straighten their obtained microscopy images, and/or to calibrate their instrument using such approach. The method could also be applied to reciprocal space (FFT) data exclusively, \emph{e.g.}, in quasi-particle interference experiments where FFTs of STM data measured at different bias voltages allows to characterize the dispersion behaviour of scattering electrons in the first Brillouin zone \cite{Yuan2019}.

More generally, we believe that this method is not restricted to SPM imaging techniques and could be used in other correction scenarios; as long as the image contains periodic information resulting from a object with known periodic features, it can be straightened using the present method. One can imagine an example scenario where a photograph of an object containing a periodic structure (\emph{e.g.}, textile sample) taken at an angle with respect to its normal axis (thus leading to an apparent linear distortion, supposing a long focal length) could be straightened using the method described in this paper.

\section{Statements}

\subsection{Methods and Code Availability}
The graphite sample was purchased commercially (SP-1 grade HOPG, from SP) cleaved in air and immediately inserted in the ultra-high vacuum (UHV) chamber, outgassed for $\sim 2$~h to about $450^\circ$C to get rid of adsorbed contaminants, and imaged with STM at room temperature using an Omicron VT-AFM with Pt/Ir tips. The SiO$_2$/Si grating calibration sample was purchased from NT-MDT. AFM imaging was performed using a Terra AFM \cite{Pawlowski2014} instrument. An example python script with detailed and commented steps for drift correction is freely available on GitHub at {\url{https://github.com/maximelester/driftcorrection}}.

\subsection{Author Contributions and Interest Statement}
M.L.S. designed the project, model, and code; wrote the manuscript. S.P. conducted the AFM experiment. P.J. and I.L. prepared the graphite sample and obtained the STM image, assisted with the development of the model, concept and writing of the manuscript. The authors declare no competing financial interests.

\subsection{Acknowledgments}
This work was supported by the National Science Center, Poland (M.L.S.: 2022/47/D/ST3/03216; I.L.: 2019/35/B/ST5/03956; P.J.K.: 2018/31/B/ST3/02450) and IDUB 6/JRR/2021 (M.L.S).\\

\bibliography{driftcorrection}

\end{document}




\title{Fourier Transform-Based Post-Processing Drift Compensation and Calibration Method for Scanning Probe Microscopy\\ ~ \\ {\small Supplementary Information}}%


\author{M. Le Ster}
\email{maxime.lester@fis.uni.lodz.pl}

\author{S. Paw{\l}owski}

\author{I. Lutsyk}

\author{P. J. Kowalczyk}
\email{pawel.kowalczyk@uni.lodz.pl}

\affiliation{University of Lodz, Faculty of Physics and Applied Informatics, Department of Solid-State Physics, Pomorska 149/153, Lodz, 90-236, Poland}

\date{\today}

\maketitle
\newpage

\section{General matrix $(\mathbf{a},\mathbf{b})\rightarrow(\mathbf{A},\mathbf{B})$}
The transformation matrix $M$ to turn $(\mathbf{a},\mathbf{b})$ into $(\mathbf{A},\mathbf{B})$ is
\begin{equation}
\label{eq:generalMatrix}
M = \frac{1}{a_xb_y - a_yb_x}\begin{pmatrix}
A_xb_y-B_xa_y & B_xa_x-A_xb_x\\
A_yb_y-B_ya_y & B_ya_x-A_yb_x \end{pmatrix}
\end{equation}
We label the prefactor denominator as $\Delta$.
\section{Real and reciprocal base vectors}
We define the primitive real vectors $\mathbf{r}_1$ and $\mathbf{r}_2$ as follows:
\begin{equation}
\begin{split}
\mathbf{r}_1 &= r_1 \begin{pmatrix}
\cos\theta\\
\sin\theta
\end{pmatrix}\\
\mathbf{r}_2 &= r_2 \begin{pmatrix}
\cos(\theta+\omega)\\
\sin(\theta+\omega)
\end{pmatrix}.
\end{split}
\end{equation}
The primitive reciprocal vectors are:
\begin{equation}
\begin{split}
\mathbf{k}_1 &= \frac{1}{r_1\sin\omega}\begin{pmatrix}
\sin(\theta+\omega)\\
-\cos(\theta+\omega)
\end{pmatrix}\\
\mathbf{k}_2 &= \frac{1}{r_2\sin\omega}\begin{pmatrix}
-\sin\theta\\
\cos\theta
\end{pmatrix}
\end{split}
\end{equation}
In the following, the vectors with a prime symbol correspond to the \emph{observed} vectors (\emph{i.e.}, distorted) and those without are the ideal vectors (in absence of drift, or simply after drift compensation).
\pagebreak
\section{$M_k$ matrix}
The $M_k$ matrix $(\mathbf{k}'_1, \mathbf{k}'_2)\rightarrow(\mathbf{k}_1, \mathbf{k}_2)$
 is obtained using equation (\ref{eq:generalMatrix}). First, we derive the prefactor denominator $\Delta$:
\begin{equation}
\begin{split}
\Delta &= \frac{1}{r'_1 r'_2\sin^2\omega'}\left(\cos\theta'\sin(\theta'+\omega')-
\sin\theta'\cos(\theta'+\omega')\right)\\
&= \frac{1}{r'_1 r'_2\sin\omega'}
\end{split}
\end{equation}
The $M_k$ matrix is:
{\footnotesize
\begin{equation}
\label{eq:M_k}
\begin{split}
M_k &= \frac{r'_1r'_2}{\sin\omega}\begin{pmatrix}
\dfrac{\sin(\theta+\omega)\cos\theta'}{r_1 r_2'} - \dfrac{\sin\theta\cos(\theta'+\omega')}{r'_1 r_2} &
\dfrac{\sin(\theta+\omega)\sin\theta'}{r_1 r_2'} - \dfrac{\sin\theta\sin(\theta'+\omega')}{r'_1 r_2} \\
\dfrac{\cos\theta\cos(\theta'+\omega')}{r'_1 r_2} - \dfrac{\cos(\theta+\omega)\cos\theta'}{r_1 r'_2} &
\dfrac{\cos\theta\sin(\theta'+\omega')}{r'_1 r_2} - \dfrac{\cos(\theta+\omega)\sin\theta'}{r_1 r'_2} \\
\end{pmatrix}\\
&= \frac{r'_1r'_2}{\sin\omega}\begin{pmatrix}
\dfrac{r'_1r_2\sin(\theta+\omega)\cos\theta' - r_1r'_2\sin\theta\cos(\theta'+\omega')}{r_1 r'_1 r_2 r'_2} &
\dfrac{r'_1r_2\sin(\theta+\omega)\sin\theta' - r_1r'_2\sin\theta\sin(\theta'+\omega')}{r_1 r'_1 r_2 r'_2} \\
\dfrac{r_1r'_2\cos\theta\cos(\theta'+\omega') - r'_1r_2\cos(\theta+\omega)\cos\theta'}{r_1 r'_1 r_2 r'_2} &
\dfrac{r_1 r'_2\cos\theta\sin(\theta'+\omega') - r'_1 r_2\cos(\theta+\omega)\sin\theta'}{r_1 r'_1 r_2 r'_2} \\
\end{pmatrix}\\
&= \frac{1}{r_1 r_2\sin\omega}\begin{pmatrix}
r'_1r_2\sin(\theta+\omega)\cos\theta' - r_1r'_2\sin\theta\cos(\theta'+\omega') &
r'_1r_2\sin(\theta+\omega)\sin\theta' - r_1r'_2\sin\theta\sin(\theta'+\omega') \\
r_1r'_2\cos\theta\cos(\theta'+\omega') - r'_1r_2\cos(\theta+\omega)\cos\theta' &
r_1 r'_2\cos\theta\sin(\theta'+\omega') - r'_1 r_2\cos(\theta+\omega)\sin\theta' \\
\end{pmatrix}
\end{split}
\end{equation}
}
\vspace*{\fill}
\pagebreak
\section{$M_r$ matrix}
The $M_r$ matrix turns $(\mathbf{r}_1, \mathbf{r}_2)\rightarrow(\mathbf{r}'_1, \mathbf{r}'_2)$. First, the prefactor inverse $\Delta$ is:
\begin{equation}
\begin{split}
\Delta &=  r'_1 r'_2 \left( \cos\theta'\sin(\theta'+\omega') - \sin\theta' \cos(\theta'+\omega') \right)\\
&= r'_1 r'_2 \sin\omega'
\end{split}
\end{equation}
The $M_r$ matrix, using equation (\ref{eq:generalMatrix}) is:
{\footnotesize
\begin{equation}
M_r = \frac{1}{r'_1 r'_2 \sin\omega'}\begin{pmatrix}
r_1 r'_2 \cos\theta\sin(\theta'+\omega') - r'_1 r_2 \cos(\theta+\omega)\sin\theta' &
r_2 r'_1 \cos(\theta+\omega)\cos\theta' - r_1 r'_2 \cos\theta\cos(\theta'+\omega') \\
r_1 r'_2 \sin\theta\sin(\theta'+\omega') - r'_1 r_2 \sin(\theta+\omega)\sin\theta' &
r_2 r'_1 \sin(\theta+\omega)\cos\theta' - r_1 r'_2 \sin\theta\cos(\theta'+\omega') \\
\end{pmatrix}
\end{equation}
}
We now inverse $M_r$. First we recall the $(2\times2)$ inverse formula (here with a prefactor):
\begin{equation}
\label{eq:inverse}
\begin{split}
(\lambda M)^{-1} &= \left(\lambda \begin{pmatrix}a&b\\c&d\end{pmatrix}\right)^{-1}\\
&= \begin{pmatrix} \lambda a &\lambda b\\ \lambda c & \lambda d \end{pmatrix}^{-1}\\
&= \frac{1}{\lambda^2(ad-bc)}\begin{pmatrix}\lambda d&-\lambda b\\ -\lambda c &\lambda a\end{pmatrix}\\
&= \frac{1}{\lambda}\frac{1}{ad-bc}\begin{pmatrix}d&-b\\-c&a\end{pmatrix}
\end{split}
\end{equation}
Let's now calculate the $(ad-bc)$ denominator in eq. (\ref{eq:inverse}). We have:
\begin{equation}
\begin{split}
ad-bc &= [r_1 r'_2 \cos\theta\sin(\theta'+\omega') - r'_1 r_2 \cos(\theta+\omega)\sin\theta'][r_2 r'_1 \sin(\theta+\omega)\cos\theta' - r_1 r'_2 \sin\theta\cos(\theta'+\omega')]\\
&- [r_2 r'_1 \cos(\theta+\omega)\cos\theta' - r_1 r'_2 \cos\theta\cos(\theta'+\omega')][r_1 r'_2 \sin\theta\sin(\theta'+\omega') - r'_1 r_2 \sin(\theta+\omega)\sin\theta']\\ \\
\end{split}
\end{equation}
\begin{equation}
\begin{split}
ad-bc&= r_1 r_2 r'_1 r'_2 \cos\theta\sin(\theta'+\omega')  \sin(\theta+\omega)\cos\theta'\\
&-(r_1 r'_2)^2 \cos\theta\sin(\theta'+\omega')\sin\theta\cos(\theta'+\omega')\\
&-(r'_1 r_2)^2 \cos(\theta+\omega)\sin\theta' \sin(\theta+\omega)\cos\theta'\\
&+r_1 r_2 r'_1 r'_2 \cos(\theta+\omega)\sin\theta' \sin\theta\cos(\theta'+\omega')\\
&-r_1 r_2 r'_1 r'_2 \cos(\theta+\omega)\cos\theta' \sin\theta\sin(\theta'+\omega')\\
&+(r_2 r'_1)^2 \cos(\theta+\omega)\cos\theta' \sin(\theta+\omega)\sin\theta'\\
&+(r_1 r'_2)^2 \cos\theta\cos(\theta'+\omega')\sin\theta\sin(\theta'+\omega') \\
&-r_1 r_2 r'_1 r'_2 \cos\theta\cos(\theta'+\omega')\sin(\theta+\omega)\sin\theta'\\
\end{split}
\end{equation}
All terms $(r_1r'_2)^2$ and $(r'_1r_2)^2$ cancel each other, and we have:
\begin{equation}
\label{eq:trig_init}
\begin{split}
ad-bc = r_1r'_1r_2r'_2[&\cos\theta\sin(\theta'+\omega')\sin(\theta+\omega)\cos\theta'\\
+&\cos(\theta+\omega)\sin\theta'\sin\theta\cos(\theta'+\omega')\\
-&\cos(\theta+\omega)\cos\theta'\sin\theta\sin(\theta'+\omega')\\
-&\cos\theta\cos(\theta'+\omega')\sin(\theta+\omega)\sin\theta']
\end{split}
\end{equation}
Which can be simplified to
\begin{equation}
\begin{split}
\label{eq:trig_final}
ad-bc = r_1r'_1r_2r'_2[&\cos\theta\sin(\theta+\omega)[\cos\theta'\sin(\theta'+\omega') - \sin\theta'\cos(\theta'+\omega')]\\
&+\cos(\theta+\omega)\sin\theta[\sin\theta'\cos(\theta'+\omega') - \cos\theta'\sin(\theta'+\omega')]]\\
=r_1r'_1r_2r'_2[&\cos\theta\sin(\theta+\omega)\sin\omega' - \cos(\theta+\omega)\sin\theta\sin\omega']\\
=r_1r'_1r_2r'_2&\sin\omega'[\cos\theta\sin(\theta+\omega)-\sin\theta\cos(\omega+\theta)]\\
=r_1r'_1r_2r'_2&\sin\omega'\sin\omega\\
\end{split}
\end{equation}
{\footnotesize
\begin{equation}
M_r^{-1} = \frac{r'_1r'_2\sin\omega'}{r1r'_1r_2r'_2\sin\omega'\sin\omega}\begin{pmatrix}
r_2 r'_1 \sin(\theta+\omega)\cos\theta' - r_1 r'_2 \sin\theta\cos(\theta'+\omega') &
r_1 r'_2 \cos\theta\cos(\theta'+\omega') - r_2 r'_1 \cos(\theta+\omega)\cos\theta' \\
r'_1 r_2 \sin(\theta+\omega)\sin\theta' - r_1 r'_2 \sin\theta\sin(\theta'+\omega') &
r_1 r'_2 \cos\theta\sin(\theta'+\omega') - r'_1 r_2 \cos(\theta+\omega)\sin\theta'
\end{pmatrix}
\end{equation}
}
Simplifying and taking the transpose $(M_r^{-1})^T$, we have:
{\footnotesize
\begin{equation}
(M_r^{-1})^T = \frac{1}{r_1r_2\sin\omega}\begin{pmatrix}
r_2 r'_1 \sin(\theta+\omega)\cos\theta' - r_1 r'_2 \sin\theta\cos(\theta'+\omega') &
r'_1 r_2 \sin(\theta+\omega)\sin\theta' - r_1 r'_2 \sin\theta\sin(\theta'+\omega') \\
r_1 r'_2 \cos\theta\cos(\theta'+\omega') - r_2 r'_1 \cos(\theta+\omega)\cos\theta' &
r_1 r'_2 \cos\theta\sin(\theta'+\omega') - r'_1 r_2 \cos(\theta+\omega)\sin\theta'
\end{pmatrix}
\end{equation}
}
Which is equal to $M_k$ above in equation (\ref{eq:M_k}):
{\footnotesize
\begin{equation}
M_k= \frac{1}{r_1 r_2\sin\omega}\begin{pmatrix}
r'_1r_2\sin(\theta+\omega)\cos\theta' - r_1r'_2\sin\theta\cos(\theta'+\omega') &
r'_1r_2\sin(\theta+\omega)\sin\theta' - r_1r'_2\sin\theta\sin(\theta'+\omega') \\
r_1r'_2\cos\theta\cos(\theta'+\omega') - r'_1r_2\cos(\theta+\omega)\cos\theta' &
r_1 r'_2\cos\theta\sin(\theta'+\omega') - r'_1 r_2\cos(\theta+\omega)\sin\theta' \\
\end{pmatrix}
\end{equation}
}